\documentclass[twocolumn,aps,prb,floats,floatfix,superscriptaddress]{revtex4-2}

\usepackage{amsmath}
\usepackage{graphicx}
\usepackage{dcolumn}
\usepackage[english]{babel}

\usepackage{color}
\usepackage{float}

\begin{document}

\title{Electronic structure of americium sesquioxide probed by resonant inelastic x-ray scattering}

\author{S. M. Butorin}
\affiliation{Condensed Matter Physics of Energy Materials, X-ray Photon Science, Department of Physics and Astronomy, Uppsala University, P.O. Box 516, SE-751 20 Uppsala, Sweden}
\author{D. K. Shuh}
\affiliation{Chemical Sciences Division, Lawrence Berkeley National Laboratory, MS 70A1150, One Cyclotron Road, Berkeley, CA 94720, USA}



\begin{abstract}
The Am $5d$-$5f$ resonant inelastic x-ray scattering (RIXS) data of americium sesquioxide were measured at incident photon energies throughout the Am $O_{4,5}$ edges. The experiment was supported by calculations using several model approaches. While the experimental Am $O_{4,5}$ x-ray absorption spectrum of Am$_2$O$_3$ is compared with the spectra calculated in the framework of atomic multiplet and crystal-field multiplet theories and Anderson impurity model (AIM) for the Am(III) system, the recorded Am $5d$-$5f$ RIXS data are essentially reproduced by the crystal-field multiplet calculations. A combination of the experimental scattering geometry and theoretical analysis of the character of the electronic states probed during the RIXS process confirms that the ground state of Am$_2$O$_3$ is singlet $\Gamma_1$. An appearance of the low-intense charge-transfer satellite in the Am $5d$-$5f$ RIXS spectra at an energy loss of $\sim$5.5 eV, suggests weak Am $5f$-O $2p$ hybridization which is in agreement with AIM estimations of the $5f$ occupancy from spectroscopic data in Am$_2$O$_3$ as being 6.05 electrons.
\end{abstract}


\maketitle

\section{Introduction}
The importance of americium oxides in the nuclear cycle is well recognized. Besides the intention to have them as a part of fuels for the fourth generation (GEN IV) nuclear reactors, the use of these oxides for the power sources in the deep space missions is also being exploited \cite{Wiss1,Vigier,Wiss2}. However, the experimental studies of the electronic structure of americium oxides are scarce due to their substantial radioactivity.

X-ray spectroscopy is an important tool for this type of studies and allows for measurements on very small quantities of material in question. It can provide information about the chemical state of actinides in various systems, (non)stoichiometry, oxygen/metal (O/M) ratio, local symmetry of the environment and charge distribution, as well as other parameters important for nuclear fuel performance. One of the attractive techniques is resonant inelastic x-ray scattering (RIXS) which is very sensitive to changes in chemical state and in symmetry of the environment of the studied element, in particular by probing the elementary low-energy excitations ($5f$-$5f$ excitations in case of actinide systems). The availability of high resolution allows one to monitor the distribution of the electronic states in greater detail. In contrast to commonly used x-ray absorption spectroscopy (XAS), the RIXS resolution is not limited by the core-hole lifetime broadening and depends solely on the instrumental resolution. However, a number of pioneering valence-to-core RIXS studies at U $3d$ edges of uranium compounds \cite{Butorin1,Butorin2,Zatsepin,Nordgren_sham,Kvashnina1,Kvashnina2} suffered from a limited resolution of utilized crystal-analyzers allowing one to resolve only the U $5f$-ligand $2p$ charge-transfer excitations which are mainly at higher energy losses than the $5f$-$5f$ excitations (only very recently, the high-resolution RIXS measurements at the U $3d$ edge became available \cite{Marino}).

Soon after the experimental proof \cite{Butorin3} that valence-to-core RIXS probes elementary electronic excitations ($d$-$d$ excitations in case of MnO), going to shallow edges was a natural way to improve the resolution of the RIXS technique. The $d$-$d$ RIXS measurements at the Cu $3p$ edge of the Sr$_2$CuO$_2$Cl$_2$ cuprate \cite{Kuiper} followed by the measurements at the Ni $3p$ edge of NiO (Ref. \cite{Butorin4,Chiuzbaian}) and NdNiO$_3$ (Ref. \cite{Butorin4}) which indicated the ability of the technique to provide greater details on the $d$-$d$ excitations. As to $f$-block systems, first RIXS measurements of $f$-$f$ excitations at the $4d$ edges of lanthanides were performed for the Gd oxide \cite{Moewes}, shortly followed by RIXS experiments for Ce-based heavy-fermion materials \cite{Butorin_Ce}, while the first RIXS measurements at the $5d$ edges of actinides were carried out for UF$_4$ (Ref.\cite{Butorin2}).

At actinide $5d$ thresholds, the main absorption edge is very broad due to so-called autoionization processes as it can be seen from XAS measurements (see e.g. Refs. \cite{Aono,Iwan,Fujimori,Kalkowski,Cox,Tobin,Nordgren,Butorin5,Butorin_Am}) and electron energy loss spectroscopy (EELS) \cite{Cukier,Moser,Rice,Moore,Bradley,Degueldre,Wiss1}. Such a broadening makes it difficult to determine the chemical shifts and obtain information about the ground state of the system from the XAS spectra. In this situation, the employment of the RIXS technique with its potentially unlimited resolution at the actinide $5d$ edges can be very helpful.

Although, initially pioneering $5d$-$5f$ RIXS studies at the actinide $O_{4,5}$ edges were carried out for oxides of uranium \cite{Butorin5,Butorin6,Shuh,Nordgren,Werme,Butorin7}, neptunium \cite{Werme,Butorin5,Butorin7} (see also the Supporting Information section of Ref. \cite{Butorin8}), plutonium \cite{Werme,Butorin5} and curium \cite{Kvashnina3}, the RIXS technique at the actinide $5d$ edge was only rarely used because of safety restrictions for research on radioactive samples in the ultra-high vacuum environment at sources of synchrotron radiation.

Here we present first Am $5d$-$5f$ RIXS data of Am$_2$O$_3$ recorded at a number of incident photon energies at the Am $O_{4,5}$ edges. The data are analyzed with help of the model calculations. The many-particle theoretical approaches turns out to be successful in reproducing the experimental spectra due to a significant localization of the $5f$ states in Am$_2$O$_3$.

\section{Experimental}
The Am oxide sample used for measurements was fabricated by technique used to prepare radionuclide counting plates at the Lawrence Berkeley National Laboratory (LBNL; see the "Preparation of counting sources" subsection in Ref. \cite{Moody}). The counting plate was prepared from an aqueous solution of about 2.0 mM Am-243 (better than 99.6\% Am-243 by mass) in 0.1 M HCl that was delivered by micropipette techniques to an area of $\sim$4 mm$^2$ area on a high purity Pt substrate (25.4 mm diameter). The aqueous droplets were allowed dry leaving a residue that was ring-shaped. This was followed by inductive heating to nearly 700 $^o$C under atmosphere to oxidize the material and fixing the material to the Pt substrate to preclude loss when placed in the UHV spectrometer chamber during the measurement. This process is expected to yield the Am oxide sesquioxide with an approximate composition of Am$_2$O$_3$ (Ref.\cite{Kvashnina3}). The counting plated was trimmed to 3 mm x 3 mm around the center and mounted on the sample holder with conductive tape as described below. The Am sample taken to the Advanced Light Source (ALS) was close to 1 $\mu$g of Am-243.

A specially designed sample holder, which is described in Refs. \cite{Butorin5,Smiles}, was used for the Am$_2$O$_3$ sample during the measurements. It is essentially a cylindrical can with slots for incoming and outgoing radiation. The sample is attached to the slab inside the can just behind the slot. Due to such a design, the sample holder served as a catch tray for material that might come loose during handling and the measurements, thus ensuring that no contamination will be left in the experimental chamber after the experiment.

Experiments in the energy range of the Am $O_{4,5}$ x-ray absorption edge ($5d\rightarrow5f,7p$ transitions) of Am$_2$O$_3$ were performed at beamline 7.0.1 of the ALS, LBNL employing a spherical grating monochromator \cite{Warwick}. The Am 5$d$ XAS data were measured in the total electron yield (TEY) mode using drain current on the sample. The incidence angle of the incoming photons was close to 90$^\circ$ to the surface of the sample. The monochromator resolution was set to $\sim$50 meV at 115 eV during measurements at the Am 5$d$ edge.

Am $5d$-$5f$ RIXS spectra of the sample at the Am $O_{4,5}$ x-ray absorption edge were recorded using a grazing-incidence grating spectrometer \cite{Nordgren2} with a two-dimensional detector. The 400 lines/mm grating of this spectrometer (based on the Rowland circle criterion) was employed to collect the RIXS data with the high energy resolution and the 300 lines/mm grating was used to measure the RIXS spectra in a larger energy range but with lower resolution. The incidence angle of the photon beam was 10$^\circ$ from the sample surface and the spectrometer was placed in the horizontal plane at an angle of 90$^\circ$, with respect to the incidence beam. The total energy resolution of the RIXS spectra was estimated to be $\sim$70 meV  for the 400 lines/mm grating and $\sim$160 meV for the 300 lines/mm grating, using the full-width at half maximum (FWHM) of the elastic peak. Measured RIXS spectra were normalized to the current in the ring.

\section{Computational details}

The XAS spectra at the actinide $5d$ edges were calculated using the formalism described in Ref. \cite{Ogasawara} in order to account for the 'giant resonance', highly-inhomogeneous broadening of the spectral transitions (due to differences in core-hole lifetime of various core-excited states) and Fano effect. In our calculations, in addition to the interactions for the free ion, as used in Ref. \cite{Ogasawara}, the crystal-field splittings in the $5f$ shell and the hybridization effects between actinide and oxygen states were taken into account using the Anderson impurity model (AIM) approach \cite{Anderson}.

To simplify the calculations of the Am $5d$-$5f$ RIXS map around the Am(III) $O_{4,5}$ thresholds, the crystal-field multiplet theory approach was used because the Am $5f$-O $2p$ charge-transfer effects were not expected to be significant in americium sesquioxide (see e.g. Ref. \cite{Butorin_Am}). The RIXS map was calculated using Kramers-Heisenberg equation
\begin{eqnarray}
I_{qq^\prime}(\omega,\omega^\prime) &=& \sum_{f} \Bigl| \sum_{m} { \langle f
|D_{q^\prime}| m \rangle \langle m |D_{q}| g \rangle \over
E_{g}+{\omega^\prime}-E_{m}-i \Gamma_{m} / 2 } \Bigr|^2  \nonumber  \\
 & &\times \delta (E_{g}+{\omega^\prime}-E_{f}-{\omega}),
\end{eqnarray}
where, $| g \rangle$, $| m \rangle$, and $| f \rangle$ are the ground,intermediate, and final states with energies $E_{g}$, $E_{m}$, and $E_{f}$, respectively, while $\omega^\prime$ and $\omega$ represent energies of incident and scattered photons, respectively.  $D_{q}$ is the dipole operator, $\Gamma$ stands for the intermediate state lifetime (Lorentzian FWHM) and $q$ and $q^\prime$ are polarizations of the light with respect to the quantization axis. The experimental geometry for RIXS measurements was taken into account in the calculations as described in Ref. \cite{Nakazawa}.

The required Slater integrals $F^{k}$, $G^{k}$ and $R^{k}$, spin-orbit coupling constants $\zeta$ and matrix elements were obtained with the TT-MULTIPLETS package which combines Cowan's atomic multiplet program \cite{Cowan} (based on the Hartree-Fock method with relativistic corrections) and Butler's point-group program,\cite{Butler} which were modified by Thole \cite{Thole}, as well as the charge-transfer program written by Thole and Ogasawara.

\section{Results and discussion}
\begin{figure}[!tb]
\includegraphics[width=\columnwidth]{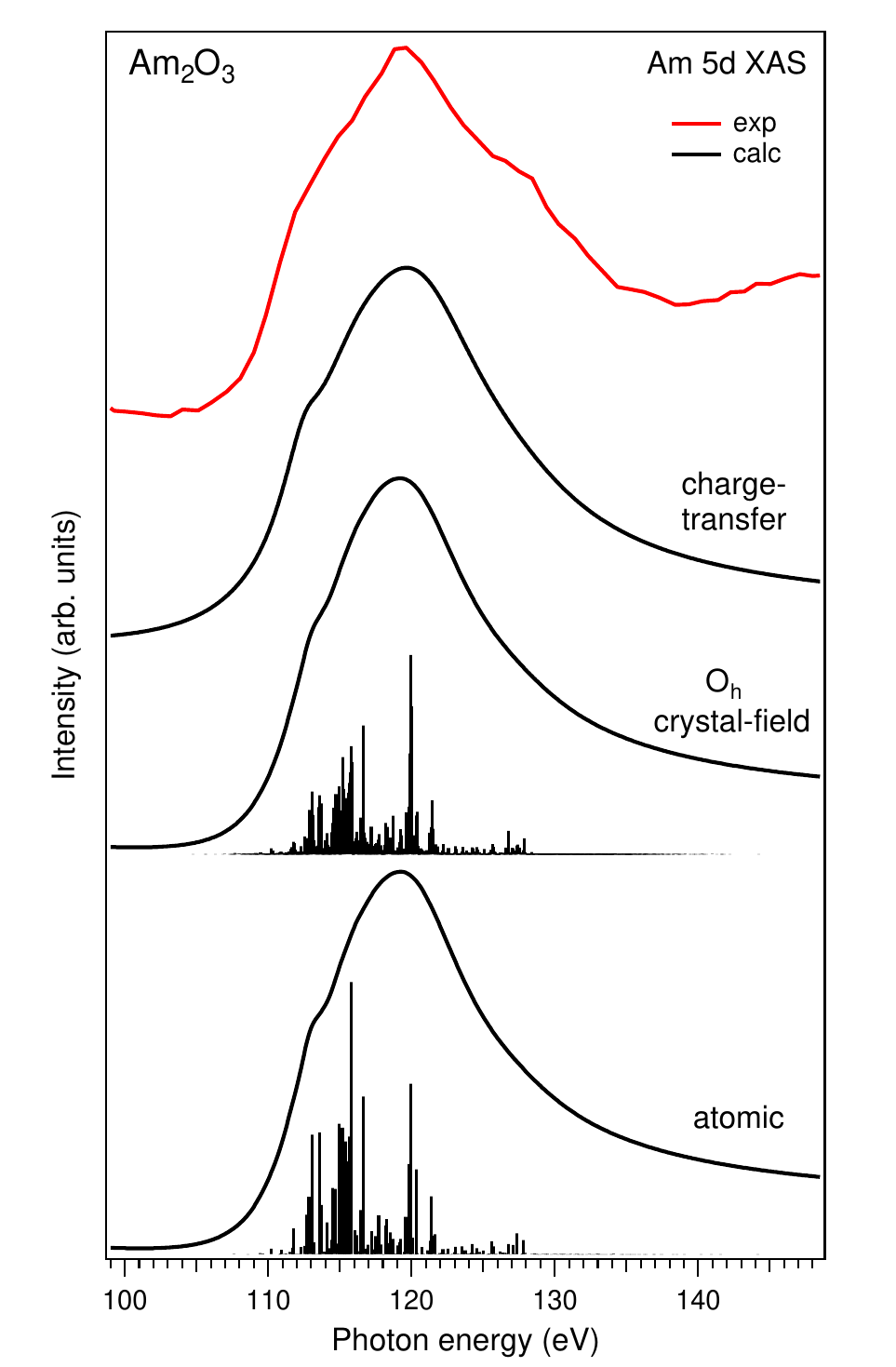}
\caption{Expeimental and calculated XAS spectra at Am $O_{4,5}$ edges of Am$_2$O$_3$. The spectra are calculated using atomic and crystal-field multiplet theory for the Am$^{3+}$ ion and Anderson impurity model, respectively. \label{5dXAS_Am2O3}}
\end{figure}

The recorded Am $O_{4,5}$ XAS spectrum of Am$_2$O$_3$ is displayed in Fig.~\ref{5dXAS_Am2O3}.
For Am$_2$O$_3$, the autoinization processes can be characterized by three decay channels following the $5d^{10}5f^{6}\rightarrow5d^{9}5f^{7}$ excitation. The $5d\rightarrow5f$ absorption and $5d\rightarrow\varepsilon{f}$ ionization are competing routes coupled to each other by the $\langle5d^{9}5f^{7}|1/r|5d^{9}5f^{6}\varepsilon{f}\rangle$ configuration interaction (CI). The decay of the excited $5d^{9}5f^{7}$ states via $\langle5d^{9}5f^{7}|1/r|5d^{9}5f^{6}\varepsilon{f}\rangle$ can be called the $5f\rightarrow\varepsilon{f}$ tunneling channel \cite{Ogasawara,Ogasawara2}. The $5d^{9}5f^{7}$ core excitations can also decay via $5d-5f5(s,p)$ Coster-Kronig and $5d-5f5f$ super-Coster-Kronig channels so that the excited states are coupled to the $5(s,p)^{-1}$ and $5f^{-1}$ ionization continua by the $\langle5d^{9}5f^{7}|1/r|5d^{10}5f^{6}5(s, p)^{-1}\varepsilon{\emph{l}}\rangle$ and $\langle5d^{9}5f^{7}|1/r|5d^{10}5f^{5}\varepsilon{\emph{l}}\rangle$ CI processes, respectively. The broadening of the $5f$ edge which is highly inhomogeneous can be evaluated by calculating these CI matrix elements.

Fig.~\ref{5dXAS_Am2O3} shows the calculated Am $O_{4,5}$ XAS spectra of Am$_2$O$_3$ compared with experimental one. First, the XAS spectrum was progressively calculated using atomic and crystal-field multiplet theory for the ($5d^{10}5f^{6}\rightarrow5d^{9}5f^{7}$) excitation of the Am(III) ion. Only the super-Coster-Kronig decay channel ($5d^{10}5f^{6}\rightarrow5d^{9}5f^{7}\rightarrow5d^{9}5f^{5}\varepsilon{\emph{l}}$) was taken into account in the calculations as a dominating process \cite{Ogasawara,Ogasawara2}.

It has been established (see e.g. \cite{Sugar,Lynch,Ogasawara3}) that the Slater integrals describing the $f$-$f$ interaction as well as the interaction of the $f$ electrons with core holes in shallow levels such as $4d$ for lanthanides and $5d$ for actinides need to be scaled down from their \emph{ab-initio} Hartree-Fock values for a description of the XAS spectra at those levels. In our calculations Slater integrals $F^{k}(5f,5f)$, $F^{k}(5d,5f)$, $G^{k}(5f,5f)$ and $R^{k}(5d5f,5d\varepsilon{\emph{l}})$ were reduced to 75\%, 75\%, 65\% and 80\%, respectively (see Table 1). The actual reduction values were also a result of the optimization of the agreement between calculated and measured Am $5d$-$5f$ RIXS spectra of Am$_2$O$_3$ (see below). Furthermore, the value of the spin-orbit coupling constant for the $5f$ shell was reduced in the calculations to 93\% from the Hartree-Fock value for the same reason.

\begin{table}
\caption{The \textit{ab-initio} Hartree-Fock values of Slater integrals and spin-orbit coupling constants for Am(III) ion. In XAS and RIXS calculations these values were reduced as described in the text. $F$, $G$ and $\zeta$ are given in eV, $R$ in $\sqrt{eV}$.}
\begin{tabular}{lc}
Am(III)&Value\\
\hline
$F^{2}(5f,5f)$&10.086\\
$F^{4}(5f,5f)$&6.577\\
$F^{6}(5f,5f)$&4.823\\
$\zeta(5f)$&0.345\\
\hline
$F^{2}(5f,5f)$&10.270\\
$F^{4}(5f,5f)$&6.707\\
$F^{6}(5f,5f)$&4.923\\
$\zeta(5f)$&0.359\\
$F^{2}(5d,5f)$&11.434\\
$F^{4}(5d,5f)$&7.377\\
$G^{1}(5d,5f)$&13.599\\
$G^{3}(5d,5f)$&8.412\\
$G^{5}(5d,5f)$&6.005\\
$\zeta(5d)$&4.002\\
$R^{1}(5d\varepsilon{g},5f^2)$&1.208\\
$R^{3}(5d\varepsilon{g},5f^2)$&0.761\\
$R^{5}(5d\varepsilon{g},5f^2)$&0.540\\
\end{tabular}
\label{table1}
\end{table}

In the crystal-field multiplet calculations, Wybourne's crystal field parameters were set to $B^{4}_{0}$=-0.835 eV and $B^{6}_{0}$=0.100 eV based on the estimates for the case of Am(III) in ThO$_2$ \cite{Hubert}. In addition, the direct interatomic exchange and superexchange, treated as a magnetic field along the z-axis and acting on the spin $S$, were set to 0.001 eV to lift the degeneracy of the states.

Furthermore, the Am $O_{4,5}$ XAS spectrum of Am$_2$O$_3$ was calculated using the AIM approach (which also included the full multiplet structure) as shown in Fig.~\ref{5dXAS_Am2O3}. The hybridization of Am $5f$ states with oxygen $2p$ states was taken into account in the AIM calculations for the Am(III) system. The same AIM parameter values were applied as in Ref.~\cite{Butorin_Am} where the Am $N_{4,5}$ XAS and $4f$ x-ray photoelectron spectra of Am$_2$O$_3$ were calculated. The ground state was described as a mixture of the $5f^{6}$ and $5f^{7}\underline{\upsilon}^{1}$ configurations with Am $5f$-O $2p$ charge transfer energy $\Delta$, $5f$-$5f$ Coulomb interaction $U_{ff}$, core-hole potential $U_{fd}$ acting on the $5f$ electrons and Am $5f$-O $2p$ hybridization strength $V$ were set to 6.5, 5.7, 6.0, and 0.7 eV, respectively.

An inspection of Fig.~\ref{5dXAS_Am2O3} reveals that the calculated XAS spectra are rather similar for all the three cases, i.e. taking into account the crystal-field interaction and Am $5f$-O $2p$ charge transfer does not lead to significant differences between calculated XAS spectra. This is because the ground state of Am$_2$O$_3$ is singlet $\Gamma_1$ (notations are for $C_{4h}$ symmetry since the finite exchange field is applied along $z$-axis) and because the $O_{4,5}$ XAS spectrum is significantly broadened due to the autoionization processes. However, such a large core-hole lifetime broadening at the $5d$ edge does not affect the resolution of the $5d$-$5f$ RIXS spectra which mainly depends on the instrumental resolution in this case.

\begin{figure}[!tb]
\includegraphics[width=\columnwidth]{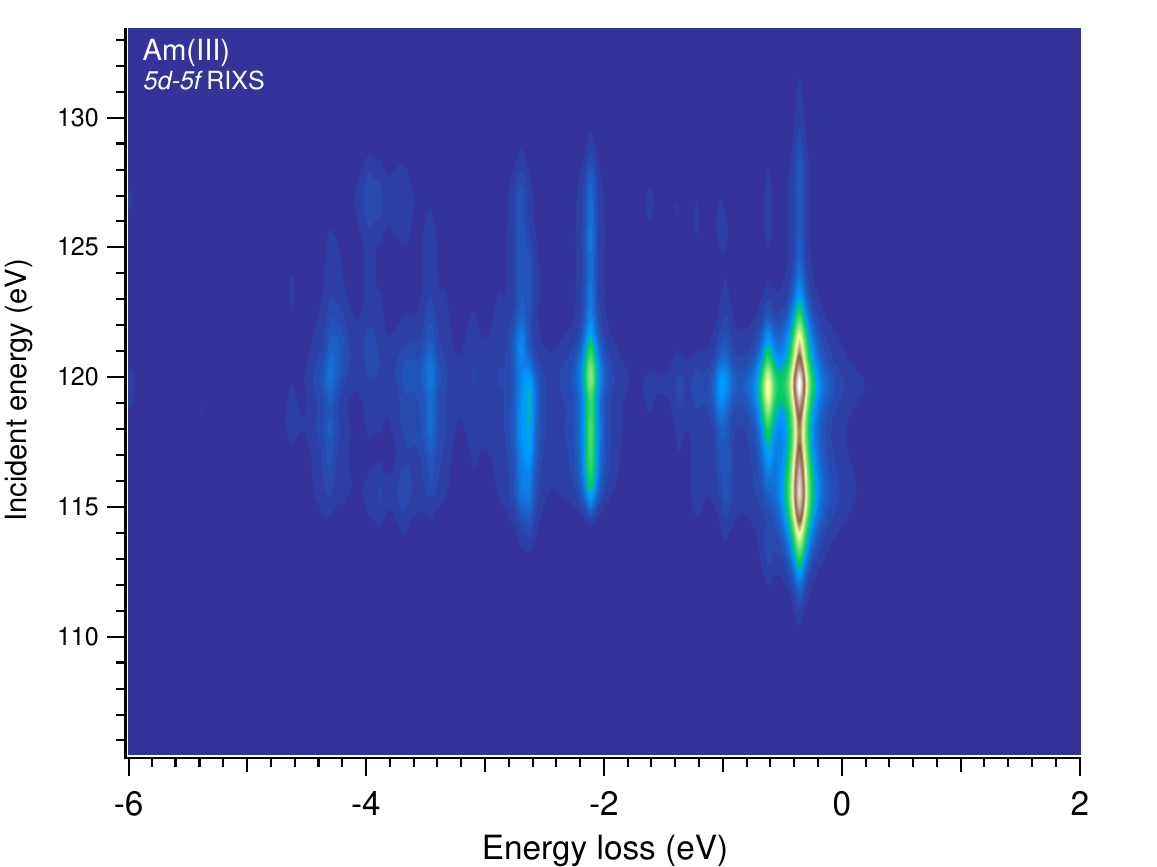}
\caption{Calculated Am $5d$-$5f$ RIXS map of Am$_2$O$_3$ using the crystal-field multiplet approach.
\label{RIXS_map_Am2O3}}
\end{figure}

To make a comparison with experimental RIXS data, we calculated the Am $5d$-$5f$ RIXS map of Am$_2$O$_3$ for incident photon energies varying throughout the Am $O_{4,5}$ edges as shown in Fig.~\ref{RIXS_map_Am2O3}. For simplicity, the crystal-field multiplet approach was used since the contribution of the Am $5f$-O $2p$ charge transfer excitations to the RIXS spectra was expected to be small. The calculated RIXS intensities are displayed using the energy-loss scale in Fig.~\ref{RIXS_map_Am2O3} and represent $5f$-$5f$ excitations. One can see that an elastic peak (at zero eV energy loss) is missing in the calculated spectra because the $5d$-$5f$ RIXS map in Fig.~\ref{RIXS_map_Am2O3} was obtained for the scattering geometry used in our RIXS experiment. For the singlet $\Gamma_1$ ground state, the $\Gamma_1\rightarrow5d^{9}5f^{7}\rightarrow\Gamma_1$ transitions are forbidden for the scattering geometry employed. In this geometry, mainly $\Gamma_3$ and $\Gamma_4$ manifold states are probed as the final states of the RIXS process. The first scattering structure which gains a non-zero intensity appears at an energy loss of 360 meV and is the most intense on the map.


\begin{figure}[!tb]
\includegraphics[width=\columnwidth]{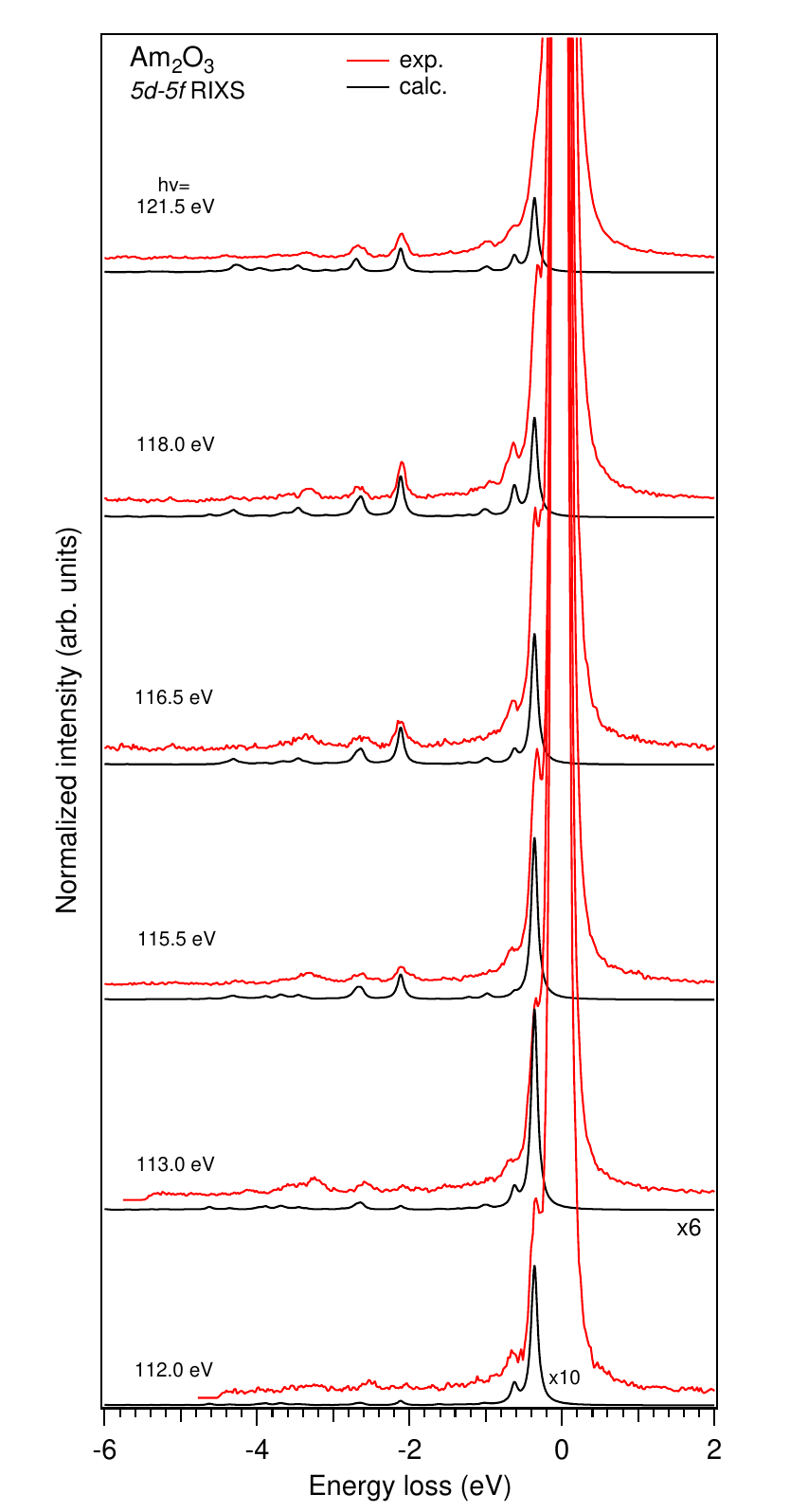}
\caption{Experimental and calculated Am $5d$-$5f$ RIXS spectra of Am$_2$O$_3$ for several incident photon energies at Am $O_{4,5}$ edges.
\label{RIXS_compare}}
\end{figure}

The experimental Am $5d$-$5f$ RIXS spectra of Am$_2$O$_3$ recorded with the 400 lines/mm grating at several incident photon energies at the Am $O_{4,5}$ edges are compared with the calculated results in Fig.~\ref{RIXS_compare}. All the experimental spectra have huge elastic peaks as a result of the reflectivity contribution in this energy range, nevertheless the first RIXS structure at $\sim$360 meV is still resolved in the recorded data. The calculated energies of the $f$-$f$ (crystal-field) excitations and relative intensities of the calculated RIXS structures are in good correspondence with the measured ones (except for RIXS structures on the slope of the elastic peak). We avoided a subtraction of the elastic peak from experimental spectra because, for the correct procedure, elastic scattering needs to be measured from the surface with exactly the same (smooth) profile but without Am$_2$O$_3$ which is difficult to achieve.

An inspection of Fig.~\ref{RIXS_compare} shows the RIXS structures are not significantly resonating with varying incident photon energies as a result of the very short core-hole lifetime in the intermediate state of the RIXS process. Some dependence on the excitation energy can be seen for the structure at an energy loss of about 3.3 eV. As a whole, in contrast to the discussion on the dual character of $5f$ states in some actinide systems, a good agreement between calculated results and experimental data in our case indicates a significant localization of the $5f$ states in Am$_2$O$_3$.

\begin{figure}[!tb]
\includegraphics[width=\columnwidth]{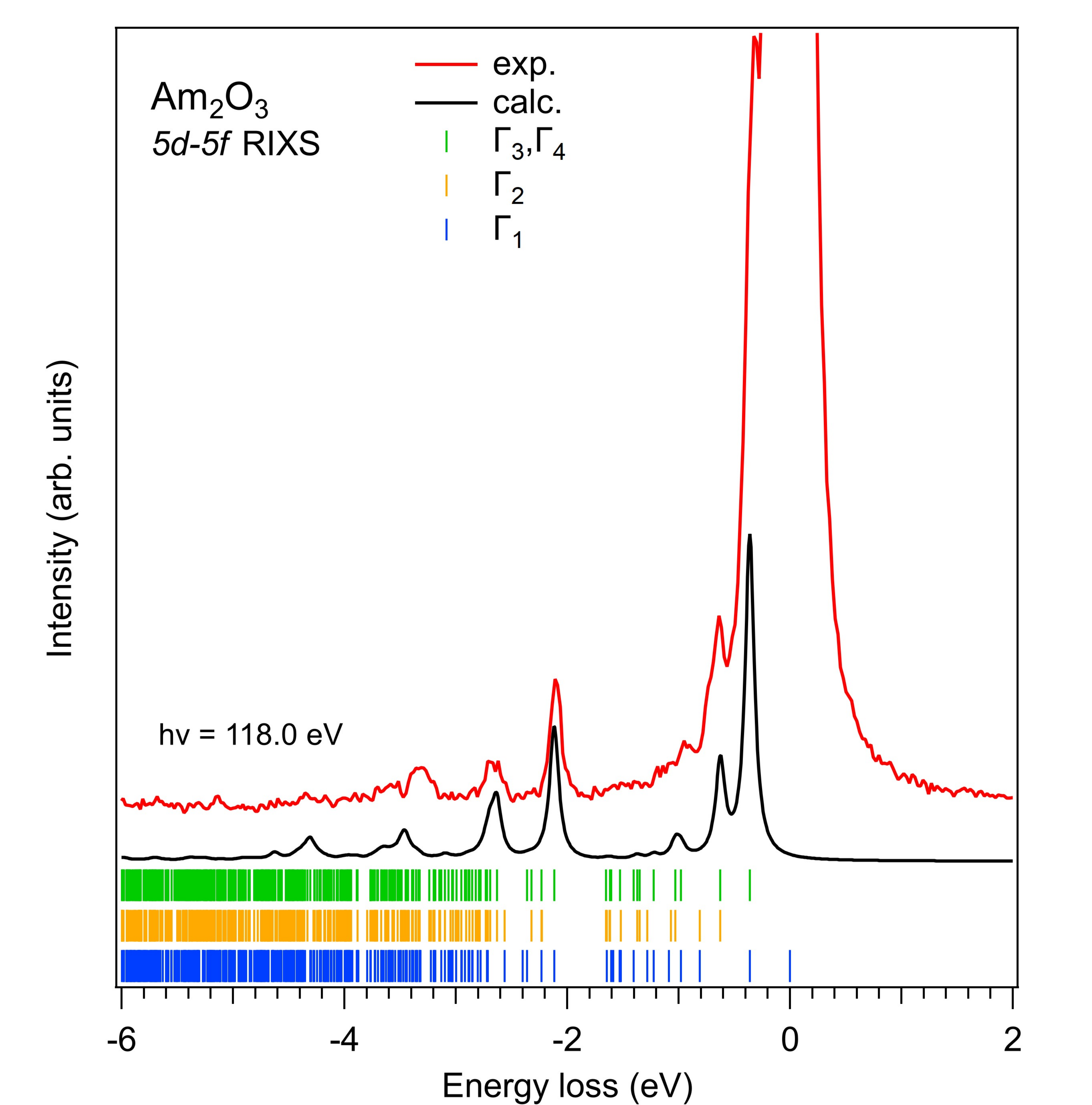}
\caption{Measured and calculated Am $5d$-$5f$ RIXS spectra of Am$_2$O$_3$ at an incident photon energy of 118.0 eV together with calculated energy diagram for $\Gamma_1$, $\Gamma_3$, $\Gamma_4$ and $\Gamma_2$ states of the $5f^6$ ground state configuration.
\label{RIXS_118}}
\end{figure}

The first RIXS structure at the energy loss of $\sim$360 meV corresponds to transitions to the $\Gamma_4$ and $\Gamma_3$ states which are split by 0.001 eV. This structure is characteristic for the Am(III) chemical state (see Fig.~\ref{RIXS_118}) and an existence of this structure in the RIXS spectra can be used to distinguish between Am(III) and Am(IV) compounds. Fig.~\ref{RIXS_118} shows experimental and calculated Am $5d$-$5f$ RIXS spectra of Am$_2$O$_3$ at the incident photon energy of 118.0 eV together with the calculated energy diagram for the $\Gamma_1$, $\Gamma_3$, $\Gamma_4$ and $\Gamma_2$ states of the $5f^6$ ground state configuration. Such a comparison of spectra with the energy diagram indicates the influence of the chosen scattering geometry on which states of the ground state configuration can be probed as a result of the RIXS process.

\begin{figure}[!tb]
\includegraphics[width=\columnwidth]{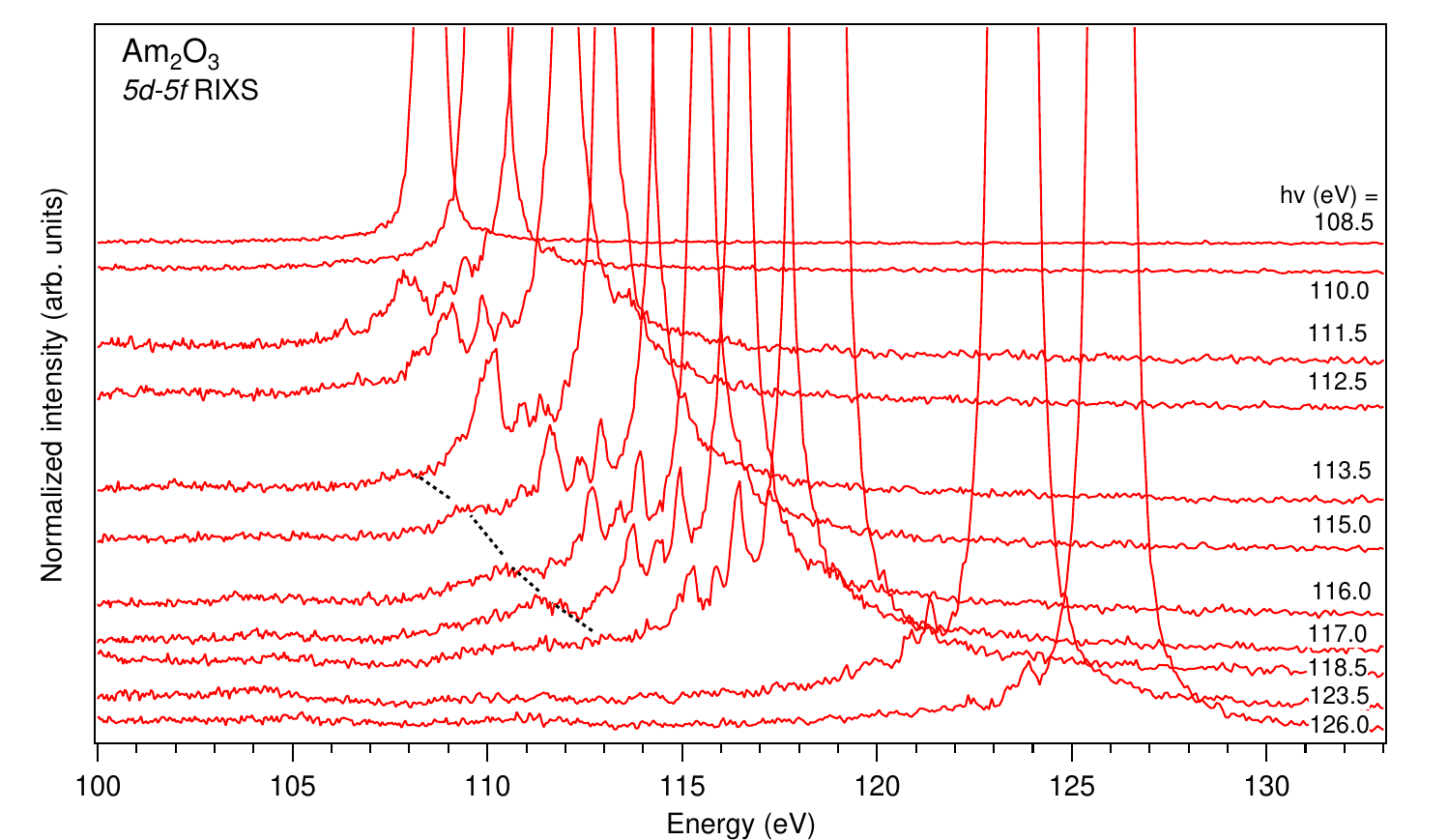}
\caption{Am $5d$-$5f$ RIXS spectra of Am$_2$O$_3$ for several incident photon energies at Am $O_{4,5}$ edges measured with lower resolution but for a more extended energy range.
\label{RIXS_3gr}}
\end{figure}

For RIXS spectra displayed in Fig.~\ref{RIXS_compare} which were recorded with the 400 lines/mm grating, the intensity is significantly cut-off for energy losses of around 5 eV and higher due to the use of the round detector (for example, a full cut-off is observed for a spectrum recorded at the incident energy of 112.0 eV). Therefore, the 300 lines/mm grating was used which provided the opportunity to measure the Am $5d$-$5f$ RIXS spectra for a more extended energy range but with lower resolution. Fig.~\ref{RIXS_3gr} displays such spectra of Am$_2$O$_3$  for a number of the excitation energies on the incident photon energy scale. The excitation energies used can be identified by present huge elastic peaks. The $\sim$360-meV-energy-loss RIXS structure is not resolved but the resonating behavior of the $\sim$3.3-eV-energy-loss RIXS structure appears to be more distinct.

In RIXS spectra recorded at excitation energies between 113.5 and 117.0 eV, a new structure at $\sim$5.5 eV below the elastic peak can be identified which behavior upon changing excitation energy is indicated by a dashed line. This structure follows the increasing excitation energy (follows the elastic peak) and can be assigned to RIXS structures. Since the crystal-field multiplet calculations do not suggest the existence of the significant RIXS structures at the energy losses higher than 4.6 eV, the new structure can be attributed to the result of the Am $5f$-O $2p$ charge transfer (charge transfer satellite). For the description of the ground state as a mixture of the $5f^{6}$ and $5f^{7}\underline{\upsilon}^{1}$ configurations, the RIXS transitions to the anti-bonding combination of the $\Gamma_1$ character is also forbidden in the scattering geometry employed, while some transitions to the weakly bonded states of the $5f^{7}\underline{\upsilon}^{1}$ configuration are allowed (see also discussion in Ref.~\cite{Nakazawa} related to RIXS spectra of CeO$_2$). These latter transitions are a result of an involvement of the O $2p$ states which are rather close to the valence band maximum. The observed 5.5-eV energy loss of the charge-transfer satellite is consistent with the choice of the value of Am $5f$-O $2p$ charge-transfer energy $\Delta$=6.5 eV as a parameter in our AIM calculations (see also Ref.~\cite{Butorin_Am}), because the $\Delta$ value is related to the center of the O $2p$ band.

\section{Conclusions}
The obtained Am $5d$-$5f$ RIXS data helps characterize the electronic structure of Am$_2$O$_3$. The observed agreement between recorded RIXS spectra and calculated results using the crystal-field multiplet approach indicates the significantly localized character of the Am $5f$ states in this oxide. The analysis of the RIXS structures by comparing the experimental and calculated data confirms that the ground state of Am$_2$O$_3$ is singlet $\Gamma_1$ of the $5f^6$ configuration. The experimental scattering geometry used, which leads to restrictions for specific RIXS transitions, made the analysis easier, despite the presence of the intense elastic peak due to high reflectivity contribution. The observation of the weak charge-transfer satellite in the Am $5d$-$5f$ RIXS spectra at the energy loss of $\sim$5.5 eV is consistent with Am $5f$-O $2p$ charge-transfer energy $\Delta$=6.5 eV used in the AIM calculations of XAS and x-ray photoemission spectra of Am$_2$O$_3$ (Ref.~\cite{Butorin_Am}), estimating the $5f$ occupancy to be 6.05 electrons in the ground state.

\section{Acknowledgments}
S.M.B acknowledges the support from the Swedish Research Council (research grant 2017-06465). The computations and data handling were enabled by resources provided by the Swedish National Infrastructure for Computing (SNIC) at National Supercomputer Centre at Link\"{o}ping University partially funded by the Swedish Research Council through grant agreement no. 2018-05973.

This work was supported in part by the Director, Office of Science, Office of Basic Energy Sciences, Division of Chemical Sciences, Geosciences, and Biosciences Heavy Elements Chemistry program (DKS) of the U.S. Department of Energy at Lawrence Berkeley National Laboratory under Contract No. DE-AC02-05CH11231. This research used resources of the Advanced Light Source, a U.S. DOE Office of Science User Facility under contract No. DE-AC02-05CH11231. The Am-243 used in this work was supplied by the U.S. DOE through the transplutonium element production facilities at ORNL.




\bibliography{An_5d_edge}
\end{document}